\documentclass[preprint,unsortedaddress,amsmath,amssymb,superscriptaddress]{revtex4-1}
\usepackage{epsf}
\usepackage{graphicx}
\usepackage{color}

\usepackage{graphicx}
\usepackage{bm}

\usepackage{newfloat}

\usepackage{hyperref}
\hypersetup{
    colorlinks=true, %set true if you want colored links
    linktoc=all,     %set to all if you want both sections and subsections linked
    linkcolor=blue,
    citecolor=blue,
    urlcolor=black,
    breaklinks=true
    }  %choose some color if you want links to stand out
\usepackage[all]{hypcap}%hyperref point to top of figure

\renewcommand{\baselinestretch}{1.3}

\begin{document}
%\linenumbers

%\title{Screening effects of the metallic outer CuO$_2$ plane enhaning the pair coherence in the inner CuO$_2$ planes  in the multilayered high-$T_{\rm c}$ cuprate supercondutors}

\title{Superconducting coherence boosted by outer-layer metallic screening in multilayered cuprates}

\author{Junhyeok~Jeong}
\affiliation{ISSP, University of Tokyo, Kashiwa, Chiba 277-8581, Japan}

 \author{Kifu~Kurokawa}
 \affiliation{ISSP, University of Tokyo, Kashiwa, Chiba 277-8581, Japan}
% \author{Yamato~Enomoto}
% \affiliation{Department of Applied Electronics, Tokyo University of Science, Tokyo 125-8585, Japan}

 \author{Shiro~Sakai} 
\affiliation{Faculty of Science and Technology, Sophia University, Tokyo 102-8554, Japan}
% \author{Yoshimitsu~Kohama}
% \affiliation{ISSP, University of Tokyo, Kashiwa, Chiba 277-8581, Japan}

 \author{Tomotaka~Nakayama}
 \affiliation{Department of Applied Electronics, Tokyo University of Science, Tokyo 125-8585, Japan}

 \author{Kotaro~Ando}
 \affiliation{Department of Applied Electronics, Tokyo University of Science, Tokyo 125-8585, Japan}

 \author{Naoshi~Ogane}
 \affiliation{Department of Applied Electronics, Tokyo University of Science, Tokyo 125-8585, Japan}

 \author{Soonsang~Huh}
 \affiliation{ISSP, University of Tokyo, Kashiwa, Chiba 277-8581, Japan}

\author{Matthew~D.~Watson}
\affiliation{Diamond Light Source Ltd, Harwell Science and Innovation Campus, Didcot, OX110DE, U.K.}

\author{Timur~K.~Kim}
\affiliation{Diamond Light Source Ltd, Harwell Science and Innovation Campus, Didcot, OX110DE, U.K.}
   
\author{Cephise~Cacho}
 \affiliation{Diamond Light Source Ltd, Harwell Science and Innovation Campus, Didcot, OX110DE, U.K.}

\author{Chun~Lin}
\affiliation{Stanford Synchrotron Radiation Lightsource, SLAC National Accelerator Laboratory, 2575 Sand Hill Road, Menlo Park, CA 94025, U.S.A.}
 
  \author{Makoto~Hashimoto}
 \affiliation{Stanford Synchrotron Radiation Lightsource, SLAC National Accelerator Laboratory, 2575 Sand Hill Road, Menlo Park, CA 94025, U.S.A.}
 
 \author{Donghui~Lu}
 \affiliation{Stanford Synchrotron Radiation Lightsource, SLAC National Accelerator Laboratory, 2575 Sand Hill Road, Menlo Park, CA 94025, U.S.A.}

 \author{Takami~Tohyama}
 \affiliation{Department of Applied Physics, Tokyo University of Science, Tokyo 125-8585, Japan}

\author{Kazuyasu~Tokiwa} 
\email{tokiwa@rs.tus.ac.jp}
\affiliation{Department of Applied Electronics, Tokyo University of Science, Tokyo 125-8585, Japan}

\author{Takeshi~Kondo}
\email{kondo1215@issp.u-tokyo.ac.jp}
\affiliation{ISSP, University of Tokyo, Kashiwa, Chiba 277-8581, Japan}
\affiliation{Trans-scale Quantum Science Institute, The University of Tokyo, Bunkyo-ku, Tokyo 113-0033, Japan}

\date{\today}
\maketitle

%\begin{abstract}
\textbf{In multilayered high-$T_c$ cuprates with three or more CuO$_2$ layers per unit cell, the inner CuO$_2$ planes (IPs) are spatially separated from the dopant layers and thus remain cleaner than the outer planes (OPs). While both interlayer coupling and the presence of clean IPs have been proposed as key factors enhancing superconductivity, their individual roles have been difficult to disentangle, as IPs and OPs typically become superconducting simultaneously. Here we investigate five-layer (Cu,C)Ba$_2$Ca$_4$Cu$_5$O$_y$ (Cu1245) with $T_{\rm c}$ = 78~K and three-layer Ba$_2$Ca$_2$Cu$_3$O$_6$(F,O)$_2$ (F0223) with $T_{\rm c}$ = 100~K using angle-resolved photoemission spectroscopy (ARPES) for the first time, and uncover an unprecedented situation, in which only the IPs become superconducting while the OPs remain metallic at low temperatures. Model calculations indicate that more than 95$\%$ of the OP wavefunction remains confined to OP itself, with minimal hybridization from the superconducting IPs, due to a large energy (or doping) difference between OPs and IPs.
In particular, we experimentally realize an ideal configuration: a single superconducting CuO$_2$ layer sandwiched between heavily overdoped metallic outer layers, which screen disorder originating from the dopant layers. 
Strikingly, this clean CuO$_2$ layer exhibits the largest superconducting gap among all known cuprates ($\Delta_0$$\sim$60~meV) and coherent Bogoliubov peaks extending beyond the antiferromagnetic zone boundary---long regarded as the boundary beyond which coherence vanishes in heavily underdoped cuprates. 
Furthermore, a widely extended coherent flat band emerges at the Brillouin zone edge, overcoming the pseudogap damping effect. Our results introduce a new physical parameter, ``the degree of screening,'' to investigate the competition between superconductivity and the pseudogap, potentially shedding new light on the longstanding debate regarding its origin---whether a charge density wave (CDW), a pair density wave (PDW), or other mechanisms. The naturally realized, nearly disorder-free superconducting CuO$_2$ layers offer a model platform for bridging the gap between typically disordered real materials and idealized theoretical models, which generally neglect disorder effects. This will thereby open a new route toward resolving one of the central challenges in condensed matter physics: the microscopic mechanism of high-temperature superconductivity.
}
%\end{abstract}

\begin{center}
  \textbf{I. Introduction}
\end{center}

Understanding the microscopic origin of high-temperature superconductivity in cuprates is arguably one of the most central unsolved problems in condensed matter physics.
To date, most investigations have concentrated on single- and double-layered cuprates, due to their structural simplicity and relative ease of crystal growth. However, scanning tunneling microscopy~\cite{Pan_STM,McElroy2005} has revealed that CuO$_2$ layers in these systems are inevitably disordered because of their close proximity to dopant layers, which introduce random potentials.
Such disorder effects complicate direct comparison with microscopic theories, which generally assume ideal, disorder-free CuO$_2$ sheets. Resolving this theory-reality mismatch may bring a breakthrough in formulating a microscopic theory of pairing in high-$T_{\rm c}$ cuprate research. 

This fundamental problem can be overcome by the multilayered cuprates, in which three or more CuO$_2$ planes stack within a single unit cell. In general notation, the CuO$_2$ planes in the multilayered system are classified into inner planes (IPs) and outer planes (OPs) with their different doping levels (or local potentials) depending on the relative distance to dopant layers. The OPs directly contact the dopant layers, whereas the IPs do not and are protected by adjacent OPs. The elemental-site sensitive nuclear magnetic resonance (NMR) measurement demonstrates that the IPs are flat and clean, maintaining a homogeneous electronic state, which is unattainable in single- and double-layer compounds~\cite{Shimizu2012,Mukuda2012}. Above all, we should keep in mind that three-layer cuprates, which achieve the highest superconducting (SC) transition temperatures ($T_{\rm c}$) among existing materials~\cite{Iyo2007}, also host clean IPs; thereby, unveiling the electronic properties of IPs and their relationship with OPs holds the key for elucidating the pairing mechanism in the high-$T_{\rm c}$ cuprates.

Recently, five- and six-layer cuprates with apical-fluorine structures have been found to exhibit small Fermi pockets in clean inner CuO$_2$ layers, which is a long-anticipated electronic structure that had eluded experimental realization \cite{Kunisada2020,Kurokawa2023}. Interestingly, while superconductivity occurs in the second inner and outer layers, the innermost layer is reported to remain metallic at the lowest temperatures. However, the SC gap observed in these systems was considerably smaller than that in other cuprates \cite{Kondo,Zhong2018,Vishik2012,zhong2022Lifshitz,Nakayama2007,Ideta2012,Kunisada2017,Vishik2014}, leaving open the possibility that the innermost layer hosts proximity-induced superconductivity, which is too weak to detect. 
In addition, their underdoped outer layer hosts superconductivity with low carrier density, making it highly susceptible to disorder introduced by the directly adjacent dopant layers. Such a disordered OP environment may hinder the penetration of superconducting proximity effects into the inner layers.

A better understanding of multilayer cuprates may be gained by exploring the inverse configuration:
The inner, cleaner layer is superconducting, but the outer layers remain metallic at the lowest temperatures. It is particularly intriguing if one could realize a special situation where the IPs open the largest SC gap among all cuprates reported to date, and the OPs are metallic with highly dense carriers. This naturally realized rare heterostructure offers a unique opportunity to distinguish intertwined effects across multiple layers on enhancing $T_{\rm c}$ in cuprates. Most importantly, this rather unique but simple material setting, where CuO$_2$ layers are shielded by metallic layers with dense mobile carriers, may be able to fill a serious gap between real materials and idealized theoretical models, where disorder effects are usually neglected. Moreover, this specific material system may provide insights into artificial heterostructures combining high-$T_{\rm c}$ cuprates and other compounds, such as topological insulators, with the ultimate goal of realizing topological superconductivity \cite{Veldhorst2012,Yang2012,Zareapour2012,Wang2013}. This study, therefore, not only allows investigating the properties of multilayer cuprates but also sheds light on the interlayer physics. 

In this article, we report an unprecedented configuration in cuprates, where superconducting CuO$_2$ planes are sandwiched between highly conductive metallic sheets. These metallic sheets effectively screen out disorder from the dopant layers, allowing us to introduce a new physical parameter---``the degree of screening''---to access the intrinsic electronic properties of cuprates. In particular, we demonstrate that the inner CuO$_2$ planes in a multilayer high-$T_c$ cuprate host the largest superconducting gap among all known cuprates, while the heavily overdoped metallic outer layers show no superconducting signals or proximity effect. Model calculations reveal that the large energy (or doping) difference between layers blocks interlayer coupling. This prevents superconducting proximity from the inner to the outer planes. Meanwhile, the highly conductive metallic outer layers shield the inner planes from disorder, significantly enhancing superconducting coherence and superfluid density. Consequently, even a single CuO$_2$ layer can exhibit both a large pairing gap and strong phase stiffness when disorder is removed. We detect a coherent flat band near the zone edge, shedding new light on the origin of the pseudogap---whether it stems from a charge density wave (CDW), a pair density wave (PDW), or other mechanisms. These findings offer a promising route to address the mismatch between real materials and idealized theoretical models that often neglect disorder effects. 

\begin{center}
 \textbf{II. Results}
\end{center}

We begin by presenting the overall electronic structure of Cu1245 with $T_{\rm c}$ = 78~K. In multilayered cuprates, carriers in the CuO$_2$ planes are distributed from the dopant layers. The doping level, therefore, decreases as the spatial distance from the dopant layers increases \cite{Kotegawa2001,Mori2002,Mukuda2012,Shimizu2012}. The five-layered Cu1245 (Fig. 1a) consists of three doping-inequivalent CuO$_2$ planes: outer planes (OP), and two inner planes (IP$_1$ and IP$_0$). 
% These doping differences create a potential gradient in the unit cell and reduce interlayer coupling (or wavefunction mixing). It yields bands that are mainly contributed by the wavefunction of each individual CuO$_2$ layer.
We observe essentially two Fermi surfaces (FSs), both hole-like and centered at $(\pi, \pi)$ (Fig. 1b). According to the above discussion, the larger FS is primarily from the OPs, while the smaller one originates mainly from the IPs. Although a slight band splitting between IP$_0$ and IP$_1$ is expected, it is too small to resolve. This may come not only from a limited equipment resolution, but also from 
unavoidable surface roughness due to the lack of a natural cleavage plane in Cu1245. 
This causes spectral broadening, hindering the identification of the band separation for IP$_0$ and IP$_1$. 
Thereby, we treat IP$_1$ and IP$_0$ together as a single ``IP" band. The Fermi momenta $k_{\rm F}$ extracted from the FS mapping are plotted in Fig.~1c, with the red and blue circles for IP and OP, respectively. From the FS volume of OP, we estimate a hole carrier concentration of $p(\mathrm{OP}) \sim 28\%$. In contrast, the IP band forms a truncated Fermi arc, within which coherent peaks are particularly sharp and clear. 
%This is typical of underdoped cuprates, preventing a reliable estimation of $p(\mathrm{IP})$ from the FS area.

Figures 1d-h display the ARPES band dispersions from the nodal to the antinodal region. 
%Because the matrix elements of OP are stronger than those of IP in the experimental setup used, the spectral intensities are more intense in the OP bands. 
The IP bands marked by red arrows show a band structure typical of cuprates with a $d$-wave SC gap. While the nodal dispersion crosses $E_{\rm F}$ with no gap, a SC gap starts opening off the node and gets larger toward the antinode. Near the antinodal region, the Bogoliubov back-bending becomes unclear since coherent peaks are strongly suppressed due to the development of the pseudogap \cite{Tanaka2006,Kondo2007}. This makes it difficult to determine the $k_{\rm F}$ from the gap minimum (dashed line in Fig.~1c). 

Strikingly, the OP band (blue arrows) shows very different behavior from the IPs. The band dispersion crosses the $E_{\rm F}$ along the entire FS. The contrasting results of OPs are more clearly demonstrated in Figs. 2c, d, where the raw EDCs at $k_{\rm F}$ and those symmetrized across $E_{\rm F}$ are plotted for the different Fermi angle $\phi$ (right-bottom inset in Fig. 2e). Interestingly, OP is gapless all along the FS. This is in stark contrast to a $d$-wave-like gap behavior typical for underdoped cuprates, observed in IP (Figs. 2a, b). 

Figure 2e summarizes the gap magnitude for both layers plotted against the $d$-wave form, $\Delta_{0}|\cos(k_{\rm x})-\cos(k_{\rm y})|/2$, where $\Delta_0$ is the extrapolated gap to the antinode. IP features a clear $d$-wave superconducting property, having a large $\Delta_{0}^{\text{IP}}\sim48$~meV. A deviation from the monotonic $d$-wave structure near the antinode is a hallmark of the two-gap states due to the presence of the pseudogap typical in underdoped cuprates~\cite{Tanaka2006,Lee2007,Kondo2007,Yoshida_PRL,Vishik2012}. On the other hand, OP shows no gap, which extends to the antinodal region. Given that the OP layer is heavily overdoped ($p$(OP)$\sim28\%$), and nearly constant peak width at the Fermi energy ($E_{\rm F}$) along wide proportions of FS (Figs. 2c, d), this no-gap feature indicates that OPs are in a normal metallic state, without a notable proximity effect induced from IPs. 
We note here that the current situation differs from Bi$_2$Sr$_2$Ca$_2$Cu$_3$O$_{10+\delta}$ (Bi2223), which exhibits significant interlayer mixing due to Bogoliubov band hybridization \cite{Kunisada2017,Ideta2021,Luo2023}: The superconducting OP band causes Bogoliubov back-bending. This results in a band crossing with the IP band, inducing significant interlayer mixing. This is not the case in the current compound since the OP band is in the metallic state at low temperatures. 

To quantify the absence of interlayer SC proximity effect, we conduct a tight-binding model calculation for the five-layer structure. We find that two factors are important for the proximity effect: the layer-dependent potential $e_l$ (depending on the distance from the dopant layers) and the interlayer hopping $V$ (Fig. 3g). Although both parameters lead to band separation in energy, their physical roles differ distinctly: $V$ yields interlayer wavefunction mixing, while $e_l$ difference between layers $\delta e$ suppresses it. A variety of {$V$, $e_l$} combinations can reproduce a given ARPES data. However, we can estimate the upper bound of $V$ from the data and demonstrate that such a $V$ value is not large enough to open a measurable energy gap in the OP band by the proximity effect from the superconducting IPs (see the Supplementary Information for more details).

To address this, we display in Fig. 3a the ARPES data of the FS and the band dispersions for Cuts 1, 3, and 5 shown in Fig.~1. Calculated results are shown alongside each panel. In both cases in Figs. 3b, c, the SC gap is set to $\Delta_{0}^{\text{IP}}\sim48$~meV for IP and zero for OP. Now, we simulate the bands with two specific combinations for parameters $V$ and $e_l$. In Fig. 3b, we set $V$=0 and choose appropriate $e_l$ for each layer to reproduce the data, as listed in the table of Fig. 3g. The $e_l$ for IP$_0$ and IP$_1$ can be particularly guided by the data of Cut 3, which clearly shows a splitting, as demonstrated by extracting the momentum distribution curve (MDC) along the blue arrow in the inset. 
In this case, the wavefunction mixing is zero simply because the interlayer hopping is set to zero. As expected, the OP spectra show no indication of gap opening (Fig. 3e). 

The upper bound of $V$ yielding the interlayer mixing can be found as follows. For this purpose, we focus on the splitting of the IP$_1$ and IP$_0$ bands, which is relatively small. Now, we set the potential difference $\delta e$ between IP$_1$ and IP$_0$ to zero, and reproduce their spectral splitting by choosing a value of $V$ (insets of Cut 3 in Fig. 3).   
We find that a $V$ value which reproduces the data lies between 0.02 and 0.04~eV. Using the larger $V$ (0.04~eV), the calculated band splitting of IPs exceeds the experimental one, as demonstrated in the inset of Fig. 3c, which plots an MDC along the blue arrow. Notably, the calculations show triple bands splitting, which is not observed in the ARPES data. 
Most importantly, even the overestimated $V$ (0.04~eV) does not open a gap in the OP band (Fig. 3f).
This indicates that the SC proximity effect from IPs to the OP band is negligible in the real material. 

In Fig. 3h, we estimate the percentage of the wavefunction confinement within OP with respect to $V$ values. We note that the OP has two low-energy eigenstates, since there are two equivalent sheets of the outer layer in the crystal structure (Fig. 1a). As a result, we get two different low-energy states for the $k_{\rm F}$ points of OP: symmetric (sym.) and antisymmetric (antisym.). Figure 3h displays both results. Even for the upper bound $V$ (0.04~eV), both results show that more than 95$\%$ of the spectral weight for the OP band is derived from the OP itself without significant wavefunction mixing from IPs. We note this OP-isolation is mainly due to a large $e_l$ difference between OP and IPs. This suggests that a large energy (or doping) difference between layers suppresses wavefunction mixing, rendering the SC proximity effect negligible in the five-layer Cu1245. 

The five-layer Cu1245 enabled us to conclude that the interlayer hopping between the heavily overdoped OP and the underdoped IP can be weak enough to render their interlayer mixing negligible. However, the presence of mutually coupled three IPs (two IP$_1$s and one IP$_0$) obscures the intrinsic properties of a clean, isolated CuO$_2$ plane. Moreover, it has been reported that the IPs in Cu1245 are more sensitive to disorder effects than those in other multilayered systems~\cite{Mukuda2006}. To avoid these complexities, we now shift our focus to the three-layer F0223. This compound corresponds to $n=3$ in $\mathrm{Ba}_2\mathrm{Ca}_{n-1}\mathrm{Cu}_n\mathrm{O}_{2n}(\mathrm{F}_y\mathrm{O}_{1-y})_2$. Note that the observation of a small Fermi pocket was achieved in the lightly doped, clean IPs for the five-layer case ($n=5$)~\cite{Kunisada2020}. The three-layer F0223 has the simplest multilayer structure with IPs, in which a single IP is sandwiched between two OPs in the unit cell (Fig. 4a). This compound is therefore ideally suited for the current study.
% This simple structure may unveil clearer features of the isolated, clean IP. 
Typically, three-layer cuprates have the highest $T_{\rm c}$ among series compositions~\cite{Iyo2007}. A widely accepted view to explain the particularly high $T_{\rm c}$ is that the underdoped IP hosts a large pairing gap but suffers from strong phase fluctuations, whereas the overdoped OP compensates for this by providing phase stiffness~\cite{Kivelson2002,Berg2008,Okamoto2008}. 
Although intriguing, and supported by experiment \cite{Kunisada2017,Ideta2021,Luo2023}, this scenario premises a hybridization between OP and IP~\cite{Berg2008,Okamoto2008}.
An intriguing question is what happens if this interlayer hybridization is negligibly small, and the IP essentially remains isolated in the three-layer structure; this will be addressed below.

Figure~4b presents the FS mapping of pristine F0223 with $T_{\rm c}$ = 100~K, showing two spectral branches of IP and OP, which lie close to each other but are clearly separated. We find that both IP and OP open the SC gap (Fig. 4e-h, and 4m), similar to those in optimally doped Bi2223: $\Delta_0^{\mathrm{IP}}$$\sim$60~meV, and $\Delta_0^{\mathrm{OP}}$$\sim$35~meV \cite{Ideta2010,Kunisada2017}. Yet, a clear difference is observed: Not only the IP band but also the OP band shows a two-gap behavior, deviating from the monotonic $d$-wave symmetry (Fig. 4m). The spectra with a larger energy scale exhibit a broad lineshape indicative of the pseudogap near the antinode (Figs. 4g, green arrows). This indicates that both IP and OP bands are located in the underdoped regime, being consistent with the lower $T_{\rm c}$=100 K of the current sample than $T_{\rm c} \sim 120$~K for optimally doped F0223~\cite{Shimizu2011_F0223,Shimizu2012}.

For both IP and OP, the FSs show arc-like features: Coherent, sharp peaks are observed only in the limited region near the node, and broad spectra characteristic of a pseudogap are observed outside that arc region \cite{Tanaka2006,Kondo2007,Shen2005}. This makes it difficult to estimate doping levels accurately from the FS areas. The momentum splitting between IP and OP is resolvable, yet is notably smaller than that in optimally doped Bi2223 \cite{Ideta2010,Kunisada2017}. This indicates that the doping-level difference between IP and OP in the present F0223 sample is smaller than that of optimally doped Bi2223. Arguments in Fig. 3 further suggest that the interlayer coupling likely exceeds that in optimally doped Bi2223.

To suppress the interlayer coupling (or band hybridization) and reproduce the situation of an isolated clean IP, we perform surface doping on F0223 and enhance the doping level difference between the IP and OP. The widely used technique for surface doping employs alkali-metal deposition~\cite{Damascelli_dosing,Zhou_dosing,Kurokawa2023}, which, however, decreases hole-doping toward the underdoped regime. For hole-doping in cuprates, \emph{in-situ} ozone annealing has been demonstrated to be effective~\cite{Drozdov2018,Zhong2018}. In this study, instead, we used a rather simple yet robust method: exposing the cleaved surface to a small air leak ($\sim3 \times 10^{-9}$ Torr) in the measurement vacuum chamber. This method effectively increases the doping level toward the overdoped side, as previously demonstrated~\cite{Palczewski2010}.

Unlike the pristine sample, the FS mapping of the surface-doped sample reveals a clearer separation between the IP and OP (Fig. 4c). In Fig. 4d, we extract momentum distribution curves (MDCs) along the nodal cuts indicated in Figs. 4b, c and fit them with Voigt functions to quantify the $k_{\rm F}$ splitting~\cite{Smit2024}. After surface doping, the separation between the IP and OP increases by more than a factor of two. 
Interestingly, however, the IP peak positions remain nearly unchanged, indicating that only the OP is significantly doped. This is further supported by full extraction of $k_{\rm F}$ points along the FS within the red box shown in Figs. 4b, c (inset of Fig. 4m). The OP band shifts significantly to the overdoped region. This lifts the saddle point at $(\pi, 0)$ above $E_{\rm F}$, causing the Lifshitz transition from the original hole-like topology centered at ($\pi,\pi$) to an electron-like topology centered at (0, 0) \cite{Kondo2004,Drozdov2018,zhong2022Lifshitz}. Consequently, the doping level ($p$) estimated from the FS area is drastically increased from the underdoped regime ($p(\text{OP})$$\sim$$8\%$) to the heavily overdoped regime ($p(\text{OP})$$\sim$$30\%$), which typically lies outside the superconducting $T_{\rm c}$ dome \cite{Keimer2015}. In contrast, the IP band remains nearly unchanged between the pristine and surface-doped samples. A small amount of carrier doping in the IPs may still occur, though the estimated amount is less than 1$\%$, as determined from the errors in the $k_F$ position. The discrepancy in doping between OPs and IPs can be naturally explained by their multilayered structure. The surface doping primarily influences the dopant layer (cleavage plane~\cite{Tokiwa}); consequently, only the OPs adjacent to it are effectively doped.

We find that the layer-selective doping also alters the layer dependence of the SC gap. 
Figures 4i-l plot the raw and symmetrized EDCs for OP and IP at various $k_{\rm F}$ points of surface-doped sample. Notably, the OP gap becomes entirely closed (Fig. 4j). This indicates that the OP becomes a heavily overdoped normal metallic state, similarly to the OPs of the five-layer Cu1245. By contrast, the IP gap remains robust and nearly identical between the pristine and surface-doped samples (Figs. 4h, l). Figure 4m presents the extracted SC gap magnitude as a function of the $d$-wave form. The extrapolated IP gap to antinode ($\Delta_{0}^{\text{IP}}$) reaches $\sim60$~meV, which is comparable to the IP gaps of other three-layered cuprates such as Bi2223 and HgBa$_2$Ca$_2$Cu$_3$O$_{8+\delta}$ (Hg1223), which has the highest $T_{\rm c}$ in cuprates ($T_{\rm c}$=130~K)~\cite{Ideta2010,Kunisada2017,Luo2023,Horio2025}. In stark contrast, the energy gap in OP, which is $\Delta_{0}^{\text{OP}}\sim35$~meV in the pristine sample, vanishes in the doped sample. Notably, both the gap magnitude and the FS topology remain unchanged in IP, indicating that surface doping has little effect on the doping level of the IP.

Importantly, despite minimized doping variation in the IP, the SC coherence peaks in the IP band become significantly sharper in the surface-doped sample (Figs. 4g, k). In contrast, the spectrum of OP becomes significantly broadened, even though it is in a heavily overdoped condition. This indicates that the sample surface is damaged during the surface doping process; however, it also demonstrates that the IP is protected from the OP, and the sharpening of the IP spectra should be intrinsic. Notably, the MDC peaks of the pristine sample are sharper than those of the surface-doped sample (Fig.~4d). The surface degradation may have reduced the spectral quality of the IP, to some extent, in the surface-doped sample. Considering such adverse conditions, this intrinsic sharpening is particularly striking.
In the heavily underdoped cuprates, it is established that the coherent peaks, observed around the nodal region, vanish across the antiferromagnetic zone boundary (AFZB) toward the antinode. This has been revealed by both ARPES~\cite{Tanaka2006,Kondo2007,Kondo,Yoshida_PRL,Chatterjee,Vishik2012} and scanning tunneling microscopy (STM) measurements~\cite{Kohsaka_Nature}. Notably, this behavior is also observed in the band of the underdoped inner CuO$_2$ plane in the optimally doped three-layered cuprates~\cite{Luo2023,Horio2025}, including Hg1223 with the highest $T_{\rm c}$ in cuprates ($T_{\rm c}$=130~K)~\cite{Horio2025}. 
This feature is also confirmed in the underdoped IP band of the current pristine sample (colored arrows in Fig.~4g). Intriguingly, however, the surface-doped IP exhibits coherent, sharp peaks in the same momentum region (colored arrows in Fig.~4l).
The enlarged Bogoliubov band (or an extended portion of the Fermi surface that contributes to superconductivity) implies a substantial increase in superfluid density; thus, the SC phase stiffness is strengthened. Thus, a large pairing gap ($\Delta_{0}^{\text{IP}}$$\sim$60~meV, the largest in cuprates) and strong phase stiffness are simultaneously realized in the single CuO$_2$ sheet. It is known that even a single-layer cuprate can reach a high $T_{\rm c}$ of $\sim$100~K, as exemplified in HgBa$_2$CuO$_{4+\delta}$ (Hg1201) \cite{Barisic2008_Hg1201}. Notably, $\Delta_0$ of Hg1201 is only $\sim40$~meV \cite{Vishik2014}, which is 1.5 times smaller than that of the IP of F0223 ($\Delta_{0}^{\text{IP}}~\sim60$~meV). Although the $T_{\rm c}$ of the surface-doped state is not determined, it opens possibility that the system with a single clean IP shielded by metallic layers with dense carriers may be capable of reaching a $T_{\rm c}$ well above 100 K. 
Note that more heavily doped samples of F0223 indeed show $T_{\rm c} \sim 120$~K~\cite{Shimizu2011_F0223,Shimizu2012}. Although they are polycrystalline, they may be in a condition similar to that of our surface-doped state. The synthesis of a single crystal with such a higher doping level has not yet been achieved; if successful, it would allow both $T_{\rm c}$ and ARPES measurements, offering a compelling direction for future research.

The simplest setting of a single CuO$_2$ layer per unit cell without disorder will provide a model case to compare the microscopic theory for the pairing mechanism, which does not include the disorder effect. We further find a rather drastic change in the IP band, particularly near the antinode, due to the SC coherence enhancement. Figures 5b-f compare the energy-momentum dispersions for both the pristine and surface‐doped crystals, taken along the momentum cuts labeled as Cuts 1 to 5 (orange lines in Fig. 5a). In the pristine sample, as one moves from Cut 1 toward Cut 5 (approaching the antinode), both the IP and OP bands become blurred and weak in intensity, reflecting the pseudogap-induced spectral damping characteristic of underdoped cuprates. 

In contrast, the IP band in the surface-doped sample remains sharp and well-defined even at the antinode (Cut 5). 
Meanwhile, the OP band shifts above $E_{\rm F}$ around ($\pi,0$), consistent with a Lifshitz transition into the heavily overdoped regime~\cite{Kondo2004,Drozdov2018,zhong2022Lifshitz}. Interestingly, the band dispersion of the coherent peak is much shallower than that before doping (for example, compare panels for Cut 3 in Fig. 5d). This indicates that the coherent peak dispersion is highly renormalized. To quantify the surface doping effect on the IP band, we integrated the EDCs over a defined momentum window (marked by grey lines in Fig. 5f) for both samples and compared them in Fig. 5g. In the pristine sample, only a hump-like feature is observed around -65 meV in the IP band. 
In the surface-doped sample, a sharp peak with almost the same width as that of the SC coherent peak near the node emerges at the same energy position as the hump in the pristine sample. Remarkably, the coherent peaks show a very flat band near the Brillouin zone edge. This is clearly observed along the momentum cut on the zone boundary (Fig. 5f). 

We found that the coherent flat band is extended two-dimensionally in a wide momentum range around $(\pi, 0)$.
Figure 5h shows the FS mapping of the surface-doped F0223.
We select four particular momentum cuts (Cuts A to D, indicated by white lines) centered at $(\pi, 0)$ and extract the energy-momentum dispersions in Figs. 5i-l. 
These clearly show coherent peaks and isotropic flatness; the coherent flat band of IP two-dimensionally extends over a wide momentum space around $(\pi, 0)$. This illustrates that SC coherent electrons at ($\pi$, 0) are spatially localized. 
Importantly, such a flat band is absent in the pristine sample, where the IP spectra contain only a broad hump reflecting the pseudogap state, a typical feature of deeply underdoped cuprates~\cite{Tanaka2006,Kondo2007,Kondo,Yoshida_PRL,Chatterjee,Vishik2012}. 
These findings suggest that spatially localized coherent electrons emerge only when disorder is effectively screened from the dopant layer---specifically, by locating the CuO$_2$ plane between metallic layers with dense mobile carriers---thereby mitigating the pseudogap-induced spectral damping. 
 These results may indicate a competition between electron coherence and the incoherent density wave, such as CDW~\cite{CDW1,CDW2,CDW3,CDW4,CDW5,Shen2005} and PDW~\cite{PDW1,PDW2,PDW3}, and its sensitivity to the disorder induced by the dopant layer. Previous studies have revealed competition through variations of the doping level in the CuO$_2$ layer \cite{Kondo,Hashimoto2015}. Our results offer a new approach to investigating competition by tuning the screening of the CuO$_2$ layer, rather than varying the doping level. Most importantly, this approach may realize an ideal system that can be directly compared with an idealized theory, excluding the disorder effect. This provides a promising avenue for identifying the origin of the pseudogap---a key issue in understanding the microscopic physics of underdoped cuprates.

\begin{center}
\textbf{III. Conclusion and discussion}
\end{center}

In conclusion, we have experimentally realized an unprecedented situation in cuprates: single CuO$_2$ layers exhibiting the largest SC gap observed to date, sandwiched between metallic sheets with dense mobile carriers that effectively screen disorder originating from the dopant layers. Notably, no SC proximity effect was detected in the metallic layers. The strategy behind our results is an interlayer $decoupling$, achieved through excess doping of the outer layers. The resulting large doping-level (or potential) difference between layers prevents orbital hybridization, thereby decoupling the superconducting inner layers from disorder from the dopant layers, either directly or via the outer CuO$_2$ planes. This system uniquely offers an experimentally accessible, nearly ideal platform for directly validating long-standing microscopic theories of pairing, previously hindered by disorder. According to our model calculations, the proximity-induced wavefunction mixing effect is minimal due to the large doping (or potential) level difference between the layers. As a result, the wavefunctions remain localized within each CuO$_2$ layer. Intriguingly, we found that the existence of heavily overdoped metallic OP significantly enhances the SC coherence in IP, overcoming the damping effect of the pseudogap states. This points to a simultaneous realization of strong pairing and strong phase stiffness in single CuO$_2$ layers. Estimating the $T_{\rm c}$ of the surface-doped state---or the $T_{\rm c}$ of single crystals under a similar condition, if successfully synthesized---will be a key future step to further validate and strengthen the present conclusions.

%Estimating the $T_{\rm c}$ of the surface-doped state or of single crystals under the same situation, if successful for the synthesis, will be a key future step that can further validate and strengthen the present conclusions.

Coherent flat bands around the Brillouin zone edge demonstrate that the spatially localized coherent electrons with a specific momentum ($\pi$, 0) emerge in the screened, clean CuO$_2$ plane. These findings may indicate the partial development of strongly bound coherent pairs among $(\pi, 0)$ electrons, competing with the pseudogap state. The results demonstrate that the purity of IPs can be tuned by the metallic conductivity of OPs,  introducing the degree of screening as a new physical parameter in cuprate research. This new research direction may help resolve the reality-theory mismatch, which has been one of the major obstacles to constructing microscopic pairing mechanisms in high-$T_{\rm c}$ cuprates.

Our results provide several further implications. In the study of artificial heterostructures, superconductors---including cuprates---are often used as a substrate source to induce SC proximity effects \cite{Wang2013,Perconte2018}. The present results illustrate that not only the SC coherence length, as frequently argued, but also the degree of orbital hybridization between the superconductor and other materials (such as topological insulators) is crucial \cite{Yilmaz2014}. Our findings also suggest an explanation for why Y-Ba-Cu-O compounds are relatively clean systems among double-layered cuprates, as revealed by the sharp peak in NMR spectra \cite{Mukuda2012}. This is mainly because the CuO$_2$ layers are sandwiched by metallic CuO chains \cite{Hussey1997_YBCO,Lee2005_YBCO}, which can effectively screen disorder effects. 

We also point out that the SC gap in Y-Ba-Cu-O emerges in the chains only at specific momentum points where the two-dimensional FS of the CuO$_2$ layers intersects with the one-dimensional FS of the CuO chains \cite{Kondo2010_YBCO}. This occurs precisely because the electrostatic potentials of the CuO$_2$ layers and CuO chains align, allowing the hybridization between them. As a result, the proximity effect from the CuO$_2$ layers to the chains is activated, leading to the emergence of a SC gap at the momentum region where the hybridized bands intersect. Our results will open a new avenue for designing disorder-resilient, high-$T_{\rm c}$ materials based on engineered electronic screening and hybridization control.

\begin{center}
  \textbf{IV. Methods}
\end{center}

The single crystals of the five-layer compound Cu1245 ($T_{\rm c}=78$~K) and the three-layer F0223 ($T_{\rm c}=100$~K) were grown by a high-pressure method, as described elsewhere \cite{watanabe_synthesis_HP}. The $T_{\rm c}$ was determined by the onset of SC diamagnetism using a dc SQUID magnetometer. ARPES experiments were carried out at the beamline I05 of Diamond Light Source and the beamline BL5-2 of Stanford Synchrotron Radiation Lightsource.  Cu1245 and pristine F0223 were \emph{in-situ} cleaved and measured under vacuum better than $8 \times 10^{-11}$ Torr. Surface doping of F0223 was achieved by exposing the cleaved crystal surface to the air leak of $\sim 3 \times 10^{-9}$ Torr in the ARPES measurement chamber~\cite{Palczewski2010}. Measurements were performed using a photon energy of $h\nu = 55$ eV at a temperature of $T \sim 7$ K. The total energy resolution was set to $\sim$12~meV.\\

%\bibliography{bibliography}

\noindent\textbf{Acknowledgements}\\
We thank Y. Tanaka for insightful discussions, T. Yamauchi and Y. Kohama for help with the magnetic susceptibility measurements.
We thank Diamond Light Source for access to beamline I05 under proposals SI36822, SI30646, SI28930, SI25416, and Stanford Synchrotron Radiation Lightsource for access to beamline BL5-2 under proposal S-XV-ST-6368A that contributed to the results presented here.
This work was supported by the JSPS KAKENHI (Grant Numbers: JP21H04439, JP23K17351, JP25H01250, and JP25H01246), the Asahi Glass Foundation, MEXT Q-LEAP (Grant No. JPMXS0118068681), The Murata Science Foundation, The Mitsubishi Foundation, and Toray Science Foundation. 

\clearpage
\newpage 

\renewcommand{\baselinestretch}{1.15}

\begin{figure}[htbp]
\includegraphics [clip,width=0.8\textwidth]{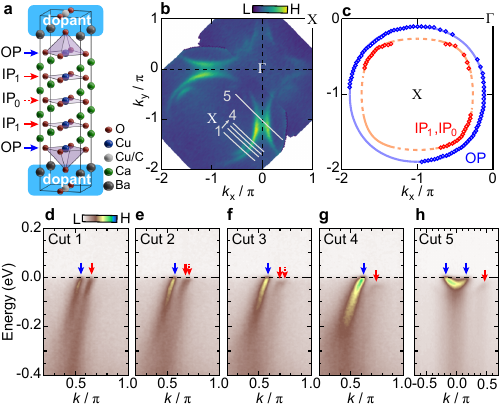}
\caption{\textbf{Fermi surface and band dispersions of the five-layer Cu1245.} 
\textbf{a}, Crystal structure of Cu1245.
\textbf{b}, Fermi surface (FS) of Cu1245 measured by ARPES. Grey lines indicate the momentum cuts used to show the dispersions in (\textbf{d}-\textbf{h}).
\textbf{c}, Fermi surface determined by the FS map in (\textbf{b}). Red and blue circles indicate the Fermi momentum ($k_{\rm F}$) points of IP and OP, respectively. Colored lines are the tight-binding fit to the data.
\textbf{d}-\textbf{h}, ARPES band dispersions along momentum cuts indicated in (\textbf{b}). Blue and red arrows mark the location of $k_{\rm F}$ for the OP and IP bands. In Cuts 2 and 3, two bands from IPs are observed. 
} 
\end{figure}

\begin{figure}[htbp]
\centering
\includegraphics [clip,width=0.7\textwidth]{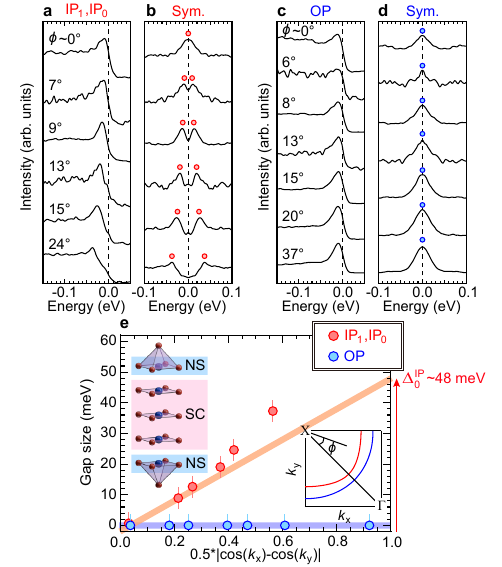}
\label{figure2}
\caption{\textbf{Superconducting gaps in the five-layer Cu1245} 
\textbf{a}, EDCs of IP taken at various $k_{\rm F}$ points. $\phi$ is the angle from the node, as indicated in the right-bottom inset of (\textbf{e}).
\textbf{b}, Symmetrized EDCs of (\textbf{a}). Colored circles indicate SC coherent peak positions.
\textbf{c}, \textbf{d}, Similar measurement as (\textbf{a}, \textbf{b}), respectively, but for the OP band.
\textbf{e}, Summary of the SC gap sizes of Cu1245, plotted as a function of $d$-wave. The top-left inset summarizes our results for the Cu1245 sample: IP is under SC state, while OP is under normal state (NS).
}
\end{figure}

\begin{figure}[htbp]
\includegraphics [clip,width=0.6\textwidth]{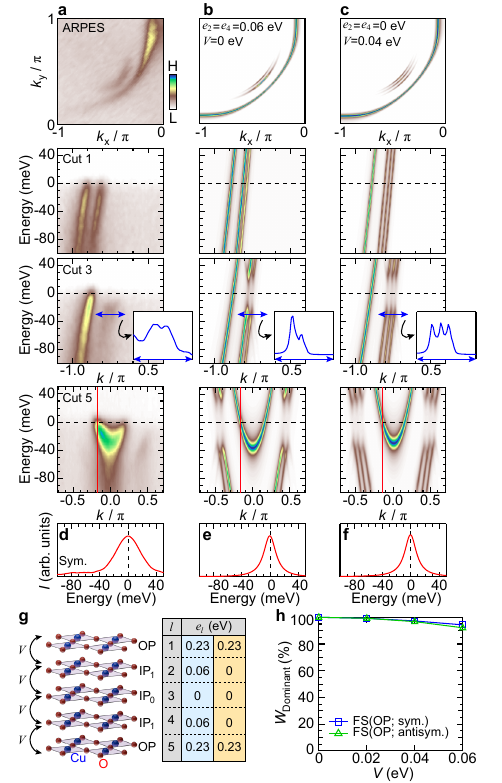}
\label{figure3}
\caption{\textbf{Model calculations for the five-layer Cu1245} 
\textbf{a}, Enlarged plots of ARPES data of Cu1245 shown in Fig. 1. The first row corresponds to FS mapping. The second, third, and fourth row correspond to Cut 1, 3, and 5 of Fig. 1(\textbf{d}), (\textbf{f}), and (\textbf{h}), respectively. The inset of Cut3 shows a MDC curve along the blue arrow. 
\textbf{b}, Model calculations based on (\textbf{a}), by setting IP$_1$ potential $e_{2}=e_{4}=0.06$~eV, and zero interlayer hopping $V=0$~eV. 
\textbf{c}, Similar model calculations with (\textbf{b}), but setting zero IP$_1$ potential $e_{2}=e_{4}=0$~eV, and interlayer hopping $V=0.04$~eV. 
For Cut 3 of (\textbf{a}-\textbf{c}), MDC along a blue arrow is extracted in the inset.
\textbf{d}, Symmetrized EDC taken at $k_{\rm F}$ from the OP band in Cut 5. This is the same curve shown in Fig. 2d ($37^{\circ}$).
\textbf{e}, \textbf{f}, Similar to (\textbf{d}), but for the model calculations of (\textbf{b}, \textbf{c}).
\textbf{g}, Schematic illustration for the model calculation. $V$ and $e_{l}$ are the interlayer hopping and layer-dependent electronic potential, respectively. Right table shows $e_{l}$ values set to obtain (\textbf{b}, \textbf{c}). 
\textbf{h}, The weight of the OP wavefunction contribution to the FS of OP itself. Since there are two equivalent OPs in the five-layer system, we obtain two different low-energy states for the FS. (sym. and antisym.). $e_{l}$ are set to the same in (\textbf{c}).
}
\end{figure}

\begin{figure*}[htbp]
\includegraphics [clip,width=0.9\textwidth]{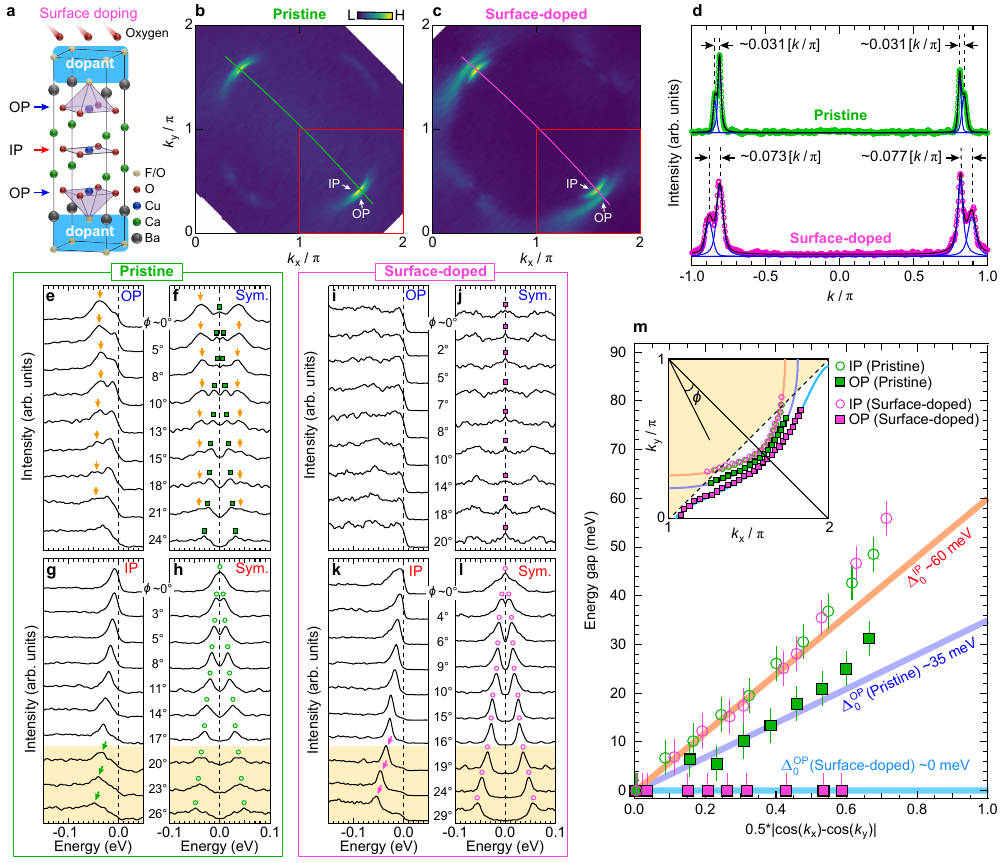}
\label{figure4}
\caption{\textbf{Fermi surface mapping and superconducting gap in the pristine and surface-doped three-layer F0223.} 
\textbf{a}, Crystal structure of F0223. We control the doping level of the sample by exposing it to a small air leak, which is known to introduce oxygen to the sample surface and dope hole carriers.
\textbf{b}, \textbf{c}, FS mappings for the pristine and the surface-doped sample, respectively.
\textbf{d}, MDC curves of the pristine and the surface-doped sample, taken along the colored line in (\textbf{b}, \textbf{c}), respectively. Black lines are fits obtained by linear backgrounds and the Voigt functions (blue lines).
\textbf{e}, \textbf{f}, Raw and symmetrized EDCs taken at various $k_{\rm F}$ points of the OP band in the pristine sample, respectively. The peaks indicated by orange arrows originate from the IP spectra. Colored squares indicate SC coherent peak positions.
\textbf{g}, \textbf{h}, Raw and symmetrized EDCs taken at various $k_{\rm F}$ points of the IP band in the pristine sample, respectively. Colored circles indicate SC coherent peak positions.
\textbf{i}, \textbf{j}, Similar to (\textbf{e}, \textbf{f}), but for the surface-doped sample.
\textbf{k}, \textbf{l}, Similar to (\textbf{g}, \textbf{h}), but for the surface-doped sample.
Colored arrows in (\textbf{g}, \textbf{k}) mark EDCs at the $k_{\rm F}$ points beyond the antiferromagnetic zone boundary (AFZB) marked by yellow shade in the inset of (\textbf{m}) toward the antinode.
\textbf{m}, Summary of the IP and OP gap magnitudes in the pristine and surface-doped sample, plotted as a function of $d$-wave form. Colored lines represent fits to the data near the node and are used to extrapolate the SC gap magnitude at the antinode ($\Delta_0$). The inset shows the determined Fermi surfaces and their fit to the tight-binding fit (colored lines). Errors of $k_{\rm F}$ is smaller than the symbol size. 
}
\end{figure*}

\begin{figure*}[htbp]
\includegraphics [clip,width=1\textwidth]{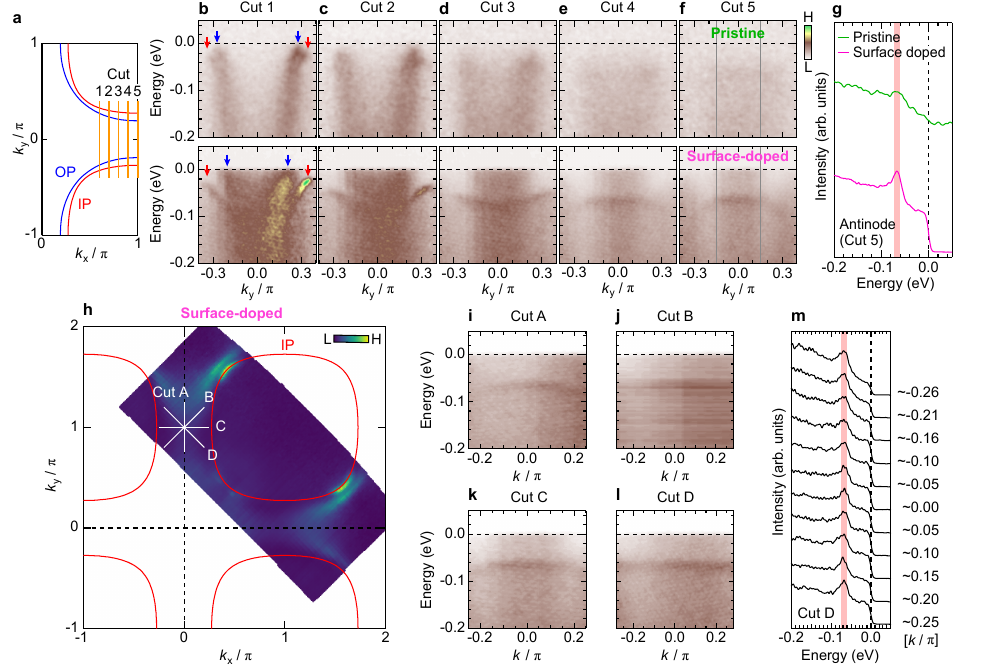}
\label{figure5}
\caption{\textbf{Emergence of widely extended coherent flat band in the surface-doped three-layer F0223.} 
\textbf{a}, Schematic FS of three-layered F0223. Orange lines indicate the momentum cuts used for (\textbf{b}-\textbf{f}).
\textbf{b}-\textbf{f}, Comparison of the band dispersions of the pristine (upper row) and the surface-doped (bottom row) sample, along the momentum cuts indicated in (\textbf{a}). IP and OP bands are marked by red and blue arrows in (\textbf{b}), respectively.
\textbf{g}, Integrated EDCs of the pristine and the surface-doped F0223 at the antinode. The integrated momentum window is chosen to [-0.15, +0.15]~$k_{y}/\pi$ (grey lines in (\textbf{f})). Each curve is normalized by the energy window of [-0.2, 0]~eV.
\textbf{h}, Fermi surface of the surface-doped F0223. Red line indicates a FS contour of IP estimated by the tight-binding method.
\textbf{i}-\textbf{l}, ARPES band dispersions obtained from the momentum cut indicated in (\textbf{h}) (Cut A to D; white lines). 
\textbf{m}, EDCs, corresponding to panel (\textbf{l}). Coherent peaks are marked by a red line, exhibiting a flat band.}
\end{figure*}

%{\bf Data availability}\\
%The data that support the findings of this study are available from the corresponding authors upon reasonable request.\\

%% author statements
%{\bf  Author Contribution}\\
%T.K. conceived and designed the project. 
%J.J. performed the ARPES experiments with the help from K.K., S.H, M.D.W., T.K.K., C.C., D.L., M.H., S.Sh., and T.K. J.J. analyzed the data with the help from K.K. and T.K. Y.E., T.N., K.A., and K.T. grew the crystals, and J.J., Y.E., T.N., K.A., and K.T. conducted the sample characterization. 
%J.J., and Y.K. performed the quantum oscillations experiments, and J.J. analyzed the data with the help from Y.K. 
%J.J., S.Sa., T.T., K.T., and T.K. interpreted the data. 
%All authors discussed the results, and J.J. and T.K. wrote the manuscript. 
%T.K. and K.T. supervised the overall project.\\

% J.J.  : Junhyeok Jeong
% K.K.  : Kifu Kurokawa
% S.H.  : Soonsang Huh
% Y.K.  : Yoshimitsu Kohama 
% S.Sa. : Shiro Sakai
% Y.E.  : Yamato Enomoto 
% T.N. : Tomoki Nakayama
% K.A. : Kotaro Ando
% M.D.W.  : Matthew D. Watson
% T.K.K.  : Timur K. Kim
% C.C.  : Cephise Cacho
% D.L.  : Donghui Lu
% M.H.  : Makoto Hashimoto
% S.Sh.  : Shik Shin
% T.T.  : Takami Tohyama
% K.T.  : Kazuyasu Tokiwa
% T.K.  : Takeshi Kondo

%{\bf  Author Information}\\
%Correspondence and requests for materials should be addressed to K.T.~(email: tokiwa@rs.tus.ac.jp) or T.K.~(email: kondo1215@issp.u-tokyo.ac.jp).\\

%{\bf  Competing Interests}\\
%The authors declare no competing interests.\\

\end{document}